\documentclass[12pt,preprint,epsfig]{aastex}

\usepackage{graphicx}
\usepackage{dcolumn}
\usepackage{bm}

\begin{document}
\title{Non-thermal transient sources from rotating black holes}
\author{Maurice H.P.M. van Putten}
\affil{Le Studium IAS, 3D Avenue Recherche Scientifique, 45071 Orl\'eans Cedex 2, France}
\author{Alok C. Gupta}
\affil{Aryabhatta Research Institute of Observational Sciences (ARIES), Manora Peak, Nainital - 263129, India}

\begin{abstract}
Rotating black holes can power the most extreme non-thermal transient sources. They have a long-duration 
viscous time-scale of spin-down and produce non-thermal emissions along their spin-axis, powered by a 
relativistic capillary effect. We report on the discovery of exponential decay in BATSE light curves of long 
GRBs by matched filtering, consistent with a viscous time-scale, and identify UHECRs energies about the GZK 
threshold in linear acceleration of ion contaminants along the black hole spin-axis, consistent with black 
hole masses and lifetimes of FR II AGN. We explain the absence of UHECRs from BL Lac objects due to UHECR 
emissions preferably at appreciable angles away from the black hole spin-axis. Black hole spin may be key 
to unification of GRBs and their host environments, and to AGN and their host galaxies. Our model points to 
long duration bursts in radio from long GRBs without supernovae and gravitational-waves from all long GRBs. 
\end{abstract}

\maketitle

\section{Introduction}

Recent developments in high-energy observations reveal a transient universe abundant in non-thermal
emissions across an exceptional range in energies, from radio, as in a recent report on an 
extragalactic $<5$ ms radio-burst \citep{lor07}, to gamma-rays in cosmological GRBs, and ultra-high
energy cosmic rays (UHECRs) at energies of $10^{19}$ eV \citep{abr07}.

Attributing non-thermal emissions and outflows to black holes is attractive, as it provides an ideal 
site for converting gravitational potential energy into various emission channels, by dissipation in 
an accretion disk or interactions with the spin of the black hole. 
Non-Newtonian behavior of the radiation processes poses novel observational challenges which may 
invite novel methods of data-analysis and novel probes of the ``physics inside" by neutrino 
emissions or gravitational-waves \citep{asp08}.

Here, we focus on rotating black holes as a potentially common engines to the most energetic
non-thermal transient sources. They are described by their mass and angular momentum (generally 
with a modest electric charge in a state of equilibrium). The outcome may depend on black hole
spin and whether accretion disks are sufficiently magnetized to create magnetic outflows. The 
physical state of nuclei is therefore important to unification schemes of AGN and quasars 
\citep{ant93,urr95,jac99} in understanding radio-morphologies, continuum and line-emissions and,
possibly, variability. The Fanaroff-Riley I and II radio-galaxies \cite{fan74} may serve as an 
example, possibly with distinctions in the presence or absence of a hidden quasar \cite{ant08}. 
The total luminosity alone, however, is sufficient to identify black hole spin \citep{liv99}.

The Pierre Auger Observatory provides the first angular correlation between UHECRs with nearby AGN
in the catalogue of \cite{ver06}. The AGN identified appear to trace nearby spiral galaxies, 
based on the complete HI Parkes All Sky Survey (HIPASS) \cite{ghi08}. It might suggest 
that stellar mass transients or dormant AGN \citep{lev01} be the source of the observed 
UHECRs. However, this does not account for the paucity of UHECRs in the Virgo cluster \citep{zaw08} 
or the apparent absence of UHECRs with blazars \citep{har07}. Classification of the associated AGN 
remains tentative, because the VCV catalogue is 
not complete and may not be an unbiased sample of the AGN Zoology. Based on the combined NASA/IPAC 
Extragalactic Database (NED), UHECRs appear to be associated with low-luminosity Seyfert galaxies and 
LINERs with relatively few radio galaxies \citep{mos08}. There may be a minimum bolometric luminosity 
for UHECRs to ensue, consistent with the paucity of UHECRs from the Virgo cluster \citep{zaw08}. UHECRs 
associated with radio-galaxies \citep{nag08}, may originate in large radio-lobes of FR II AGN 
\citep{fra08a,fra08b}. This would require the radio-jet to be baryon-rich, to account for the observed 
proton and heavier elements in the UHECRs. It would produce largely isotropic emissions in UHECRs inconsistent 
with the paucity of UHECRs from blazars that are hiding FR IIs, and does not account for the apparent 
association with Seyfert galaxies and LINERs. Instead, UHERs may indeed be produced by some of the galactic 
nuclei. If confirmed, ``UHECR-active" and ``UHECR-inactive" nuclei promises a significant extension to AGN 
classification and unification schemes.

In this {\em Letter}, we consider two first-principle physical properties of rotating black holes: a 
long-duration viscous time-scale of spin-down and ultra-high energy emissions induced by a relativistic 
capillary effect along the spin-axis \citep{van08b}:
\begin{itemize}
\item 
Exponential decay in the long duration evolution of GRB light curves, identified by application 
of matched-filtering to the BATSE catalogue on the premise that the inner engine of long GRBs is 
long-lived \citep{pir98}. The template used is generated by spin-down of a Kerr black hole 
interacting with high-density matter at critical magnetic stability \citep{van03}.
\item 
Creation of UHECRs in a linear accelerator powered by a relativistic capillary effect in the funnel 
along the spin axis of the black hole surrounded by an ion torus -- typical for AGN -- with energies 
that are tightly correlated to the mass of the supermassive black hole and the lifetime of the AGN. 
The latter has recently been studied in some detail for the FR II radio-galaxies \citep{ode08}.
\end{itemize}

We discuss the application to unification schemes for all GRBs -- long and short, with and without 
supernovae (the {\em Swift} event GRB 060614, \cite{del06}) -- and UHECR-active AGN. We begin with 
an introduction to the relevant physical properties of Kerr black holes, and apply these to the 
active nuclei of transient sources. We describe specific observational tests to enable a comparison 
with observations, present and in the future.

\section{Some physical properties of Kerr black holes and surrounding matter}

Rotating black holes in astrophysical environments introduce at least two unique physical properties: 
a long-duration timescale of spin-down associated with a large reservoir in spin-energy, and 
linear-acceleration to ultra-high energies along their spin axis.

Rotating black holes \citep{ker63} are described by their mass $M$, angular momentum $J$, specific 
angular momentum $a=J/M$, and spin-energy $E_{s}=\frac{1}{2}\Omega_H^2 I f_s^2,$
where $\sin\lambda=a/M$ \citep{van99} with relativistic correction factor
$0.7654<f_s=\frac{\cos(\lambda/2)}{\cos(\lambda/4)}<1$ for the angular velocity
$\Omega_H=\frac{1}{2M}\tan(\lambda/2),$ and where $I=4M^3$ denotes the moment of inertia in the 
limit of slow rotation\citep{tho86}. Thus, $E_{s}/M$ can reach 29\% ($\lambda =\pm \pi/2$), larger 
than the spin energy per unit mass in neutron star by an order of magnitude. Kerr black holes 
evolve according to the first law of thermodynamics $dM = \Omega_H dJ + T_H dS$ for changes in
spin energy, $dE_s=\Omega_H dJ$, and dissipation, $dQ=T_H dS_H$, with Bekenstein-Hawking entropy 
\citep{bek73}, $S_H = 4\pi M^2\cos^2(\lambda/2).$
In the limit of viscous spin-down with no radiation, we note that $S_H$ {\em doubles} in spin-down 
to zero from an initial state of maximal spin.

Frame-dragging creates {\em linear acceleration} by coupling of the Riemann tensor to the angular 
momentum $J$ of charged particles along open magnetic flux-tubes about the spin-axis of a black hole. 
In geometrical units, the product of the Riemann tensor (cm$^{-2}$) and angular momentum (cm$^{2}$)
is of dimension 1, which represents a force \citep{pap51}. Integration along a semi-infinite line 
to infinity produces a potential energy. In Boyer-Lindquist coordinates we have, along the spin-axis 
of the black hole \citep{van00,van05,van08b}, 
\begin{eqnarray}
 {\cal E}(r)=\int_r^\infty \mbox{Riemann}\times J ds = \omega J.
\label{EQN_E1}
\end{eqnarray}
The ensuing {\em relativistic capillary effect} extracts $e^\pm$-pairs from the environment
of the black hole, where they are created by canonical pair-cascade processes \citep{bla77}, out to
larger distance along the spin-axis \citep{van00}. A transition to a nearly force-free state 
\citep{bla77} terminates in an outgoing Alfv\'en front. Thus, the raw Faraday-induced horizon potential 
(\ref{EQN_E1}) is communicated to the outgoing Alfv\'en front. Since magnetic flux-surfaces 
{\em upstream} are stable against pair-cascade, they remain largely charge-free, and become
clean site for linear acceleration of baryonic contaminants, ionized by exposure to ambient 
UV-radiation.

In its lowest energy state the horizon flux $\Phi_{\theta}^e$ of a magnetic field with strength 
$B$ through a polar cap of half-opening angle $\theta_H$ is, adapted from \cite{wal74},
$\pi B \frac{(r_H^2+a^2)^2}{r_H^2+a^2\cos^2\theta}\sin^2\theta,$
where $r_H=2M\cos^2(\lambda/2)$ and $a=M\sin\lambda$. The full horizon flux
($\theta = \pi/2$) $\Phi_H^e  = 4\pi BM^2$ is hereby the same for maximal rates and zero-spin,
while through a polar cap $(\theta<<\pi/2)$, $\Phi_{\theta}^e \simeq 2 \pi B M^2 {\theta}^2$ 
about maximal spin (and a factor two larger at zero spin). The Faraday induced potential energy 
on a flux tube with $A_\phi = BM^2\theta^2$ at $r=r_H$ in (\ref{EQN_E1}), is
\begin{eqnarray}
{\cal E}(\theta) = e c \partial_t\Phi = e \Omega_H A_\phi 
      \simeq \frac{1}{2} e BM \theta^2 = 2.16 \times 10^{20} B_{5} M_9 \theta^2\mbox{~eV}
\label{EQN_EUH}
\end{eqnarray}
in the limit of maximal spin ($\Omega_H\simeq 1/2M$), where $\theta\le \theta_H$. 
The lifetime of spin of a supermassive black hole is largely based on dissipation $T_H\dot{S}_H$ 
in the event horizon. The average is about $\frac{1}{3}$-rd of the maximal dissipation rate, i.e.:
\begin{eqnarray}
 <\dot{Q}> = \frac{c}{3}(\Omega_H A_\phi)^2 
  = \frac{c}{12}B^2 M^2 = 5.6\times 10^{47} \left(B_5M_9\right)^2 \mbox{erg s}^{-1}
\label{EQN_D}
\end{eqnarray}
The lifetime of spin for a maximally spinning black hole hereby becomes
$T \simeq 29\% \times 2\times 10^{63} M_9 \mbox{~erg~} <\dot{Q}>^{-1},$ whereby
$B_5M_9^{1/2}T_7^{1/2}\simeq 1.04$, or
\begin{eqnarray}
B_5M_9 = 1.04 \sqrt{\frac{M_9}{T_7}}
\label{EQN_BM}
\end{eqnarray}
with $T=T_7 10^7$ yr following \cite{ode08}. With conservation of flux, $B\simeq 10^5$ G at the ISCO may 
be compared with $B\sim 10^{-2}$ G fields on sub-parsec scales in Mrk 501 \citep{osu08}. 
Scaling to stellar mass black holes surrounded by high-density matter with superstrong 
magnetic fields gives 
\begin{eqnarray}
 <\dot{Q}> = 6.9\times 10^{52} \left(\frac{B}{5\times 10^{15}\mbox{G}}\right)^2
\left(\frac{M_\odot}{7 M_\odot}\right)^2 \mbox{erg s}^{-1}
\label{EQN_D2}
\end{eqnarray}
with
\begin{eqnarray}
\left(\frac{B}{5\times 10^{15}\mbox{G}}\right)
\left(\frac{M}{7 M_\odot}\right) 
= 1.05 \sqrt{\frac{M ~20 \mbox{~s}}{7M_\odot T_{90}  }},
\label{EQN_T90}
\end{eqnarray}
where $T_{90}$ refers to the observed durations of long GRBs in seconds.

For the magnetic field strengths in (\ref{EQN_BM}) and (\ref{EQN_T90}), 
the mass surrouding the black hole can be estimated on the basis of the
stability bound for poloidally magnetized disks by \cite{van03}:
\begin{eqnarray}
\frac{\cal E_B}{{\cal E}_k}\simeq \frac{1}{15}
\end{eqnarray}
following two closely related models for the poloidal magnetic field of
energy ${\cal E}_B\simeq \frac{1}{6}B^2R^3$ supported by an inner disk of radius $R_D$,
mass $M_D$ and kinetic energy ${{\cal E}_k}$. To leading order, we have 
${\cal E}_k=\frac{GMM_D}{2R_D}$, so that for supermassive and stellar mass black holes
\begin{eqnarray}
   {M_D} \simeq 120 M_\odot\left(\frac{{\cal E}_k}{15{\cal E}_B}\right)
   \left(\frac{R_D}{6R_g}\right)^4 \left(\frac{M_9^2}{T_7}\right),
\end{eqnarray}
\begin{eqnarray}
   {M_D}\simeq 0.1 M_\odot \left(\frac{{\cal E}_k}{15{\cal E}_B}\right)
   \left(\frac{R_D}{6R_g}\right)^4 \left(\frac{M}{7M_\odot}\right)^2
   \left(\frac{~20 \mbox{~s}}{T_{90}}\right)
\end{eqnarray}
with characteristic matter densities of $7.9\times 10^{-11}$ g cm$^{-3}$ and,
respectively, $1.9\times 10^{11}$ g cm$^{-3}$ (close to the neutron drip line). The
associated Alfv\'en velocities $v_A/c=B/\sqrt{4\pi \rho c^2+B^2}$, where $c$ denotes
the velocity of light, are universal, $v_A[\mbox{AGN}]=0.1052 c$ and, respectively, 
$v_A[\mbox{GRB}]=0.1072 c$, and are mildly relativistic.

\section{GRBs from rotating black holes}

Long GRBs represent the complete life-cycle of a relativistic inner engine (e.g. \citep{pir98}). 
The recent {\em Swift} discovery of the long GRB 060614 poses the challenge to identify a common 
inner engine to long GBRs with and without supernovae.

In suspended accretion, the evolution of a rapidly spinning black hole ensues on a viscous timescale 
by spin-down against surrounding high-density matter \citep{van99,van01}. The durations of long GRBs 
are consistent with the lifetime of rapid spin of a black hole in interaction with high-density matter 
at its magnetic stability limit \citep{van03}. This applies to newly born, as in core-collapse of a 
massive stars \citep{woo93}, or in the binary merger of a neutron star with a companion black hole
or neutron star. Both scenarios produce a black hole-torus system \citep{van03}. 

We searched for the presence of a viscous time-scale in the GRB light-curves of long events in the
BATSE catalogue using matched filtering. To this end, we average normalized light curves in durations 
and count rates, following translations in time, for a best-fit against a template produced by viscous 
spin-down (Eqs.(7) in \cite{van08a}). Fig. 1 shows a comparison of the template with the normalized light 
curves of two notable low-variable, low-luminosity events \citep{rei01}. Because these single Fast Rise 
Exponential Decay (FRED)-like light events are rare, we next focus on typical, highly variable light curves 
from a list of consecutive BATSE triggers, and hence with no selection criteria. Fig. 2 shows some 
typical fits to the template.

Light curves of GRBs are remarkably diverse, and few look similar. 
Variability in GRB light curves of long events shows a most variable event GRB 990510, and 
least variable events such as GRB 970508 and GRB 980425, while variability and luminosity are 
correlated \citep{rei01}. The shortest time-scales of variability reflects intermittency 
of a long lived inner engine \citep{pir98}. For GRB-afterglow emissions produced by baryon-poor jets 
\citep{she90,fra01}, variability at intermediate and small time scales is expected from unsteady
radiation processes and modulations by orientation effects, especially for rotationally 
powered inner engines. An accretion disk or torus is in a state of forced turbulence 
\citep{van99} and may be precessing \citep{por00}, and their interactions with the central black hole
should be intermittent, but much less so due to time-variability in accretion in the suspended accretion 
state. Furthermore, the light curves are inherently scaled in energy, photon count rates and durations 
by their redshifts.

We extract an ensemble average in the form of a normalized light curve (nLC), wherein all short 
timescale fluctuations have been filtered out relative to the $T_{90}$ of each individual burst 
by averaging of the individually normalized light
curves against a fixed template. In our procedure, we translate and scale the data to fit a
template for producing stable zeroing of the bursts by translation in time followed by normalization 
to permit taking an ensemble average. It filters out variabilities in the ensemble of light curves at
sub-dominant time-scales, and is not focused on studying individual light curves.

Our focus is complementary to the mean spectral properties of GRBs \citep{fis89}, statistics of sub-bursts 
\citep{rei01,qui02}, linear temporal profiles in light curves \citep{mcb02} or modeling selected light 
curves to bright events \citep{por00,lei07}. Matched filtering is different from earlier approaches based 
on the BATSE timing signals of the starting time, $T_{50}$, $T_{90}$ and peak count rates \citep{mit96,fen97}.
Quite generally, template matching gives stability for burst-alignments and scaling, much more so than 
BATSE timing signals. On this basis, our results disprove the earlier suggestion of a linear decay in the 
average GRB light curves \citep{fen97}, which is also not seen in \cite{mit96}.

 We have verified that our nLC is stable against a choice of template. Fig. 3 shows very similar
 results obtained for the Kerr template and block-type template (Fig. 3). The matching procedure 
 is consistent, in that block- and Kerr-type templates produce a match of their FWHM, shown in the 
 lower window of Fig. 3. Matched filtering thus creates an nLC as a unique diagnostic for the
 underlying slow-time behavior, by filtering out the diversity in GRB light curves due to
 intermittencies, orientation effects and redshifts.

The exponential decay in the nLC (Fig. 4) is in good agreement with viscous spin-down of Kerr black holes
by dissipation of spin-energy in the event horizon which, to leading order, represents a {\em doubling} the 
Bekenstein-Hawking entropy 
\begin{eqnarray}
S_{H,i} = 2\pi M^2 \rightarrow S_{H,f}=4\pi M^2,
\label{EQN_S}
\end{eqnarray}
where $\lambda$ changes from $\lambda_i=\pi/2$ to $\lambda_f=0$. To next order, the energetic output 
of the black hole is in various radiation channels by surrounding matter and, to much higher order, 
in radiation along the spin-axis of the black hole. 

The GRB-afterglow arise by dissipation of the kinetic energy in ultra-relativistic baryon-poor 
jets in internal and external shocks \citep{she90}. Modeling by Poynting flux along an open magnetic 
flux-tube in the force-free limit \citep{gol69,bla77,tho86} supported by an equilibrium magnetic moment 
of the black hole surrounded by a uniformly magnetized disk or torus predicts a correlation between 
the peak energy $E_p$, the true energy in gamma-rays $E_\gamma$ and the durations $T_{90}$
as measured in the local rest-frame, given by \citep{van08b} 
\begin{eqnarray}
E_pT_{90}^{1/2} \propto E_{\gamma},
\label{EQN_C1}
\end{eqnarray}
showing a Pearson coefficient of 0.85 in the HETE-II and Swift data \citep{ama02,ghi04,ghi07}, compiled 
in Fig. 1 of \citep{van08b}. 

We attribute the correlation between variability and luminosity \citep{rei01} to the angular distribution 
of beamed outflows from open flux-tubes with a finite horizon opening angle, consistent with observations 
\citep{fra01}. Their output (luminosity density per sterradian) increases with angle between the
the line-of-sight and the spin-axis of the black hole, reaching its maximum along magnetic field lines 
suspended at a half-opening $\theta_H\sim M/R_T$ on the event horizon set by poloidal curvature
in the inner torus magnetosphere, where $R_T$ denotes the radius of the torus \citep{van03}.  The relatively
small fraction of spin-energy thus released in true energy in gamma-rays $E_\gamma$ is in good agreement with 
the observed estimate of $\sim 10^{51}$ erg \citep{fra01}. 
Conceivably, intrinsic variability in gamma-ray emissions reaches a maximum in the boundary layer with the 
surrounding baryon-rich torus winds. If so, the {\em observed} variability and luminosity depend on viewing 
angle. In the nLC, this orientation effect is effectively averaged out. 

\section{UHECRs from supermassive rotating black holes}
\label{sec:3}

The funnel within an ion torus creates a natural environment for an open magnetic flux-tube along the 
spin-axis of the black hole, supported by an equilibrium magnetic moment of the black hole when exposed 
to a surrounding magnetic fields \citep{van99}. The magnetic field may be intermittent in strength and 
sign, provided it carries a net poloidal magnetic flux.

A relativistic capillary effect induced by differential frame-dragging acting on open magnetic flux-tubes 
introduces a linear accelerator upstream of an outgoing Alfv\'en front illustrated in Fig. 5. Exposure to 
UV-radiation from an ion torus in AGN \cite{for95} allows for acceleration of ionic contaminants to UHECR 
energies correlated with the physical properties of the central black hole. Based on (\ref{EQN_EUH}) and
(\ref{EQN_BM}, we have
\begin{eqnarray}
{\cal E} = 5.6\times 10^{19} \sqrt{\frac{M_9}{T_7}} \left(\frac{\theta_H}{0.5}\right)^2 \mbox{~eV},
\label{EQN_CUH}
\end{eqnarray}
where an horizon half-opening angle $\theta_H\simeq 30^o$ is used as a fiducial value based on M87 
(\cite{jun99}; M87, however, is an FR I source). The underlying magnetic field-energy -- which is 
currently not accessible to direct observations -- is here replaced by its correlation to the lifetime 
of black hole spin, and its identification with the lifetime of FR II radio-galaxies.
The correlation (\ref{EQN_CUH}) is in good agreement with the GZK threshold energy of 
$6\times 10^{19}$ eV for observing UHECRs from nearby AGN.

UHECRs may be powerful probes of spin of supermassive black holes. In our model, the UHECRs are emitted 
anisotropically preferentially along an finite angle relative to the spin-axis of the black hole, as in 
gamma-ray emissions from stellar mass black holes discussed in the previous section. This angle may be 
appreciable, perhaps as large as 30$^o$ on the basis of the opening angle in M87. This picture is consistent 
with the absence of an UHECR association to BL Lac objects \citep{har07}, which are believed to include both 
FR I and II sources viewed close to the line of sight -- too close perhaps for producing UHECRs. 

The maximum luminosity $L_j$ of the baryon-poor jet along an open magnetic flux-tube can be estimated 
in terms of the fraction $\sim \frac{3}{8}\sin^4\theta_H$ of $<\dot{Q}>$,
\begin{eqnarray}
L_j = L_-+L_+\simeq  1.3 \times 10^{46} \left(\frac{M_9}{T_7}\right)
            \left(\frac{\theta_H}{0.5}\right)^4\mbox{~erg~s}^{-1},
\label{EQN_LJ}
\end{eqnarray}
where $L_-$ may drive dissipation in shocks downstream and $L_+$ may generate ultra-high energy emissions upstream. 
Here, $L_+$ depends on the efficiency of mediating the Faraday induced horizon potential to the outgoing Alfv\'en front: 
it reaches 100\% in the force-free limit envisioned by \cite{bla77} when $L_-\simeq0$. The luminosity in UHECRs 
depends on the density of ionic contaminants upstream, and their exposure to UV radiation. The observed time-averaged 
AGN luminosity, considered in \cite{wax04}, is here governed by the time-average 
$<\theta_H^4(t)>$, which does not provide 
a sharp constraint on peak energies in UHECRs. Geometrical considerations suggest 
$<\theta_H^4(t)>\simeq M^4<R_D^{-4}(t)>$ is set by the poloidal curvature in magnetic field 
lines associated with a time-variable $R_D(t)$, as in the Seyfert galaxy
MCG 60-30-15 \citep{tan95,iwa96}. 
Including the duty cycle, $<L_j(t)>$ can hereby
be smaller than the peak values (\ref{EQN_LJ}) by one to two orders of magnitude. 
Low-luminosity AGN hereby favor UHECR production by allowing for relatively clean sites
for particle acceleration. In light of (\ref{EQN_LJ}), it would therefore be of interest to pursue an observational
study of the true ages of Seyfert galaxies. See further \citep{van99a,far08} for related considerations on intermittencies.

In general, therefore, (\ref{EQN_LJ}) is an upper limit to a luminosity in UHECRs which may be intermittent on 
time-scales of the light-crossing time of the ion torus -- tens of years or less -- and extending for the lifetime 
of spin of the black hole, i.e.: 1-10 M yr, on the basis of observational lifetimes of FR II radio galaxies.

The same principles apply to UHECRs from spinning stellar mass black holes in GRBs. Expressed in total energy output, 
the relative contributions to the PAO observations satisfies
\begin{eqnarray}
\frac{{\cal E}[\mbox{GRB}]}{{\cal E}[\mbox{AGN}]} \simeq 1.4\times 10^{-4} 
\left(\frac{N[\mbox{AGN},<100\mbox{~Mpc}]}{500}\right)^{-1},
\end{eqnarray}
given 1 long GRB yr$^{-1}$ within 100 Mpc and scaled against a fiducial number of 500 AGN powered by black hole spin 
within this distance. GRBs appear to be sub-dominant.

\section{Conclusions and Outlook}

The findings in this paper are part of a study on durations, light curves, spectral properties 
and correlations to the mass and lifetime of black hole spin:
\begin{enumerate}
\item We identify the bi-modal distribution of durations in the BATSE catalogue 
      with hyper- and suspended accretion on slowly and rapidly spinning BHs
	  (Eqs.(7-8) and Eqn.(9), respectively, in \cite{van01}).
	  The time scale of hyperaccretion onto slowly spinning black holes is
      $\sim t_{ff}{{\cal E}_k}/{{\cal E}_B}~\simeq$ tenth's of seconds, 
	  where $t_{ff}$ denotes the free fall timescale. Black hole
	  angular velocities that are slower and faster than the angular velocity at the ISCO 
	  thus give rise to transient sources to the left and, respectively, to the right of the
	  break at $\sim 2$ s separating the two classes in the BATSE catalogue.
      Unification on the basis of black hole spin rates 
	  predicts low-luminosity X-ray afterglows also to short GRBs \citep{van01},  
	  confirmed by {\em Swift} in GRB 050709 and HETE II in GRB 050509B. This
	  unification does not require (but does not rule out) different values of the magnetic
	  field strength and the mass of the black hole.
\item We report on an exponential decay in the BATSE catalogue of light curves of long GRBs in agreement 
      with viscous spin-down of a Kerr black hole. The quality of the fit to the model template by matched 
	  filtering shows that long GRBs are to leading order a process of doubling the entropy of the
	  event horizon, expressed in (\ref{EQN_S}).
\item We describe a common inner engine to unify long events GRB 030329/SN2003h and GRB 060614 with and 
      without supernovae, produced in CC-SNe \citep{woo93,pac98} and the binary merger of a neutron star 
	  with a rapidly spinning black hole \citep{van99} or a companion neutron star (Fig. 1 in 
	  \cite{van03}).
\item Our model predicts a spectral-energy correlation (\ref{EQN_C1}) in agreement 
      with HETE II and {\em Swift} data with a Pearson coefficient of 0.85 (Fig. 1 in \cite{van08b}).
\item Our model predicts UHECRs with energies (\ref{EQN_CUH}) correlated to the mass of the central black 
      hole and the lifetimes of AGN in agreement with PAO data, and emitted anisotropically consistent
	  with the paucity of UHECRs from BL Lac objects.
\end{enumerate}
 
 For GRBs, accretion-powered models (e.g \citep{woo93, kum08} are different. They do not account for 
 GRB060614, since it was not a core-collapse event \citep{del06}. Winds from the inner disk, at typical 
 temperatures of 1-2 MeV, are too contaminated to produce the ultra-relativistic (baryon-poor) outflows 
 needed to account for the observed GRB-afterglow emissions and spectral-energy correlations. If attributed 
 to the central black hole, then its spin {\em up} by continuing accretion \citep{kum08} is at odds with the 
 decay in BATSE data (Fig. 1). 
 
 Late-time X-ray emissions on a timescale of 1-10 k s -- distinct from X-ray afterglows
 to the prompt GRB-emissions -- are observed by {\em Swift} in a number of long GRBs
 (see \cite{zha07} for a review). They can be attributed to late-time infall of the outer envelope of 
 a remnant progenitor star in core-collapse events \citep{kum08}. Late-time infall onto a black hole, 
 regardless of spin, may be expected to be luminous as matter forms high-densities in the 
 process of free fall (as in SN1987A). In the absence of a massive progenitor star, as in 
 mergers of a neutron star and a black hole or mergers of two neutron stars, we anticipate that events 
 such as GRB 060614 -- and more generally long GRBs with anomalously low-luminosity X-ray afterglow 
 emissions, identified with binary mergers in low density environments -- feature 
 no or relatively weak late-time X-ray emissions. Weak late-time X-ray emissions is expected, since the 
 break-up of the neutron star(s) in a binary merger is known to be potentially messy, leaving matter
 at relatively large radii to fall in with appreciable time delay. This may be verified observationally,
 by searching for late-time X-ray emissions in long GRBs without supernovae and GRBs -- long and short -- 
 with anomalously low X-ray afterglows to their prompt GRB emissions. 
 
 To illustrate our unification scheme, we interpret some pivotal GRBs as follows:\\
 {\bf GRB 050709, GRB 050509B} ({\em short GRBs with X-ray afterglows}). Binary mergers of a neutron star with a slowly spinning
 black hole, producing a short duration event in a state of hyperaccretion. The GRB-afterglows 
 were produced by black hole outflows, interacting with a low-density environment.\\
 {\bf GRB 030329} ({\em long GRB-supernova}). Core-collapse of a massive star, producing a long duration event in a 
 state of suspended accretion. GRB-afterglow emissions were produced by black hole outflows 
 from a rapidly spinning black hole, interacting with high-density environments typical for star-forming regions.\\
 {\bf GRB 050911} ({\em long GRB without X-ray afterglow}). The binary coalescence of a neutron star with a rapidly spinning 
 black hole or the merger of two neutron stars, giving rise to a long duration event
 in a state of suspended accretion. The absence of an X-ray afterglow is attributed to 
 the low-density environment typical for binaries.\\
 {\bf GRB 060614} {\em long GRB without supernova}). Binary coalescence of a neutron star with a rapidly spinning 
 black hole or the merger of two neutron stars, given rise to a long duration event
 in suspended accretion without a supernova. The relatively low-luminosity X-ray plateau
 is attributed to a limited amount of late-time infall of remnant matter following the
 break-up of the neutron star(s).\\
 {\bf GRB 070110} ({\em long GRB with late-time X-ray emissions}). A core-collapse of a 
  massive star GRB-supernova with late-time X-ray emissions from fall-back of matter from 
  the remnant stellar envelope \citep{kum08}.\\
 {\bf GRB 080319B} {(\em luminous and variable long GRB}). A GRB viewed along the boundary 
  of a baryon-poor jet, where the luminosity and variability is highest subject to 
  instabilities in interaction with surrounding baryon-rich disk winds.

If GRBs and the recently reported short duration radio burst (if confirmed) are related, then 
long-duration radio-bursts may exist associated with events like GRB 060614 or, more generally,
with long duration bursts with anomalously weak X-ray afterglows.

Complete optical-radio surveys of the local transient universe promise to provide a valuable catalogue 
of core-collapse supernovae, radio-loud SNe and radio-bursts. These surveys can be optimized by scanning 
along the local super-clusters, avoiding the large-area voids in between \citep{ein94}. They may be augmented 
by large neutrino detectors \citep{shi05} to provide real-time triggers of the onset of supernovae 
as in SN1987A. The Australian MOST survey illustrates surveys of radio-loud SNe, all of which are CC-SNe,
few of which will be associated with long GRBs in view of a branching ratio of a mere 0.2-0.4\% of Type Ib/c into 
GRBs \citep{van04}. Augmented by optical surveys, e.g., by Pan-STARRs (in the northern hemisphere), these surveys
could provide targets of interest to LIGO and Virgo, to study a possibly common inner engine
to Type II and Type Ib/c SNe, perhaps similar to those at work in GRBs, the latter which are expected to
have a long duration bursts in gravitational waves with negative chirp.
The CC-SNe scenario for long GRB-SNe requires a companion star in close binary orbit to ensure rapid 
rotation by tidal interaction \citep{pac98}. At present, a companion has only been identified to the 
Type II/Ib SN 1993J event \citep{mau04}, otherwise not known to have been a GRB.

The FR I and II classification expresses a dichotomy in radio-galaxies \cite{fan74}, as in 3C31 and Pic A. 
The FR II sources are lobe-dominated and edge-brightened, straight and one-sided (on the kpc scale), which 
is testimony to their power, exceptional long-term orientation stability and radiation beaming in 
relativistic outflows. It suggests that FR IIs may be associated with black hole spin, more likely so than 
the probably accretion powered FI Is \citep{bau95}. But, the FR II's may not be the only AGN harboring rapidly 
spinning black holes. Follow-up PAO statistics on UHECRs will be important in confirming the association with 
AGN, and in identifying the type of AGN, host galaxies and their orientation relative to the line-of-sight.
We suggest that spinning black holes may be hosted in Seyferts and LINERs, producing largely leptonic 
outflows and emissions in UHECRs without baryon-rich disk winds. UHECR-active AGN without FR II 
morphology may be understood by the absence of disk winds, perhaps due to weak poloidal magnetic fields. 
When radio-jets do form, polarization studies on FR I and II jets show large-scale ordered magnetic fields 
with more pronounced toroidal magnetic fields in FR II sources. Black hole spin interacts largely with the 
inner disk, through the {\em time-average} of energy-density of net poloidal magnetic flux, while UHECR
production is associated with {\em instantaneous} poloidal magnetic flux along an open flux-tube with finite 
opening angle set by a generally time-variable radius of the inner disk. The outflows will remain 
largely baryon-free in the absence of outer disk winds (cf. the X-ray jet of the Crab pulsar \citep{lu01}), 
inhibiting dissipation in shocks to re-accelerate charged particles (cf. modeling GRB-afterglows \citep{she90}).
There may therefore by anomalously weak or unseen extended radio-emissions associated with intermittent
UHECR production from clean sites for acceleration in Seyfert galaxies, where the details may be 
constraint by the lifetime of Seyfert galaxies.

{\bf Acknowledgment.} We thank the referee for a number of thoughtful comments and 
Ski Antonucci, Gilles Theureau and Isma\"el Cognard for stimulating discussions. 
This work is supported, in part, by a grant from the R\'egion Centre of France.

\begin{figure}
\centerline{\includegraphics[angle=00,scale=.80]{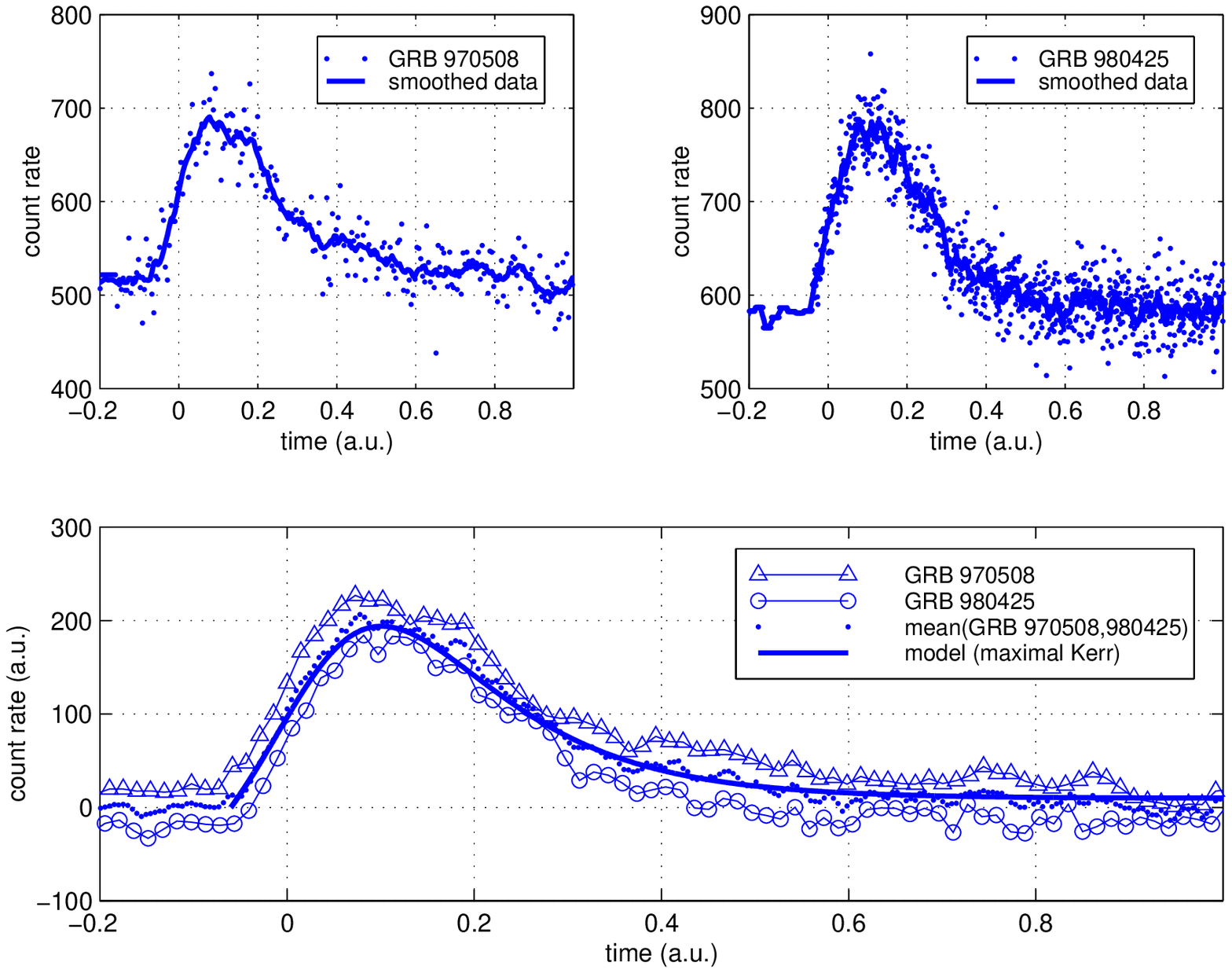}}
\caption{
  A comparison of two low-variability events GRB 970508 and GRB 980425
  and a template light curve for GRBs from rotating black holes. The two 
  events display a characteristic fast-rise and exponential decay (FRED). 
  The data are scaled in duration and count rate to compare their shape with
  the template. Note that the true time-of-onset $t_0^\prime$ is slightly 
  before the $t_0=0$ in BATSE timing.}
\end{figure}
\begin{figure}
\centerline{\includegraphics[angle=00,scale=.80]{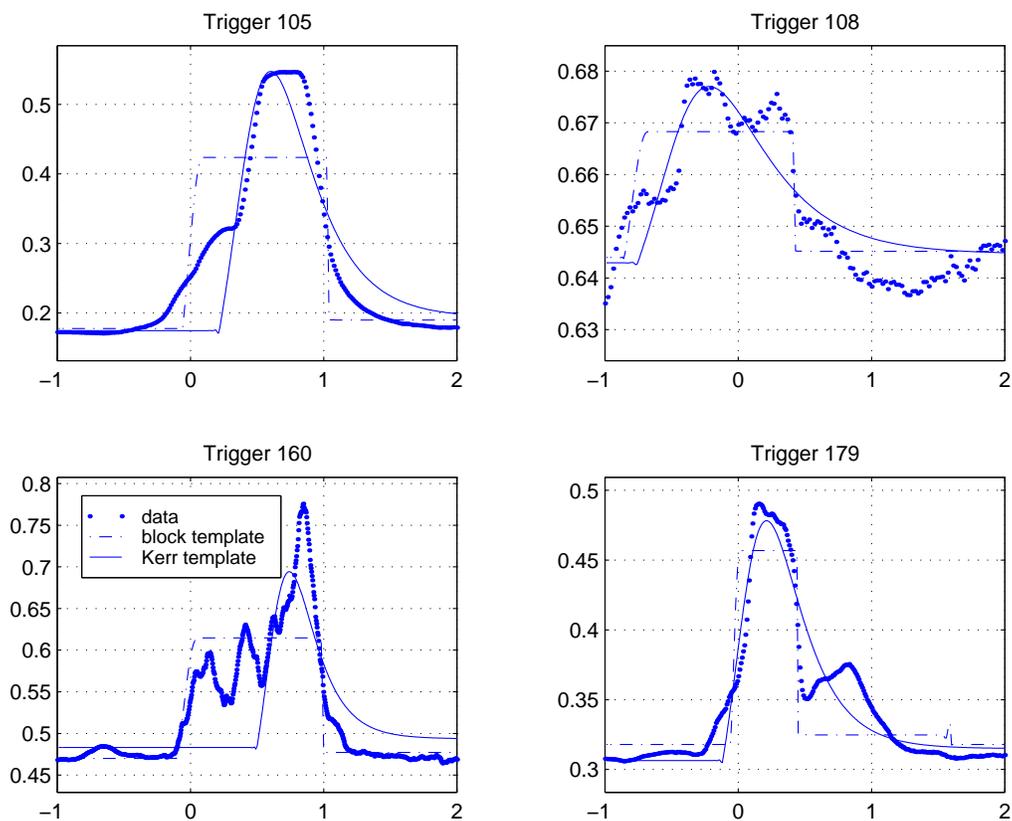}}
\caption{
  Shown are examples of template fits to various long bursts with normalized durations 
  in the sample L1: 2 s $< T_{90} <$ 20 s, for both block-type and Kerr template.
  In each window, the horizontal axis refers to normalized time (a.u.) and the 
  vertical axis refers to normalized count rates (a.u.). Fitting to a template gives
  a zeroing of the burst by translations in time defined on the basis of the entire 
  shape of the burst relative to the template, illustrated by Trigger 160, different 
  from using special instants such as the BATSE $t_0$ data. Light curves are subsequently 
  normalized and averaged to produce a normalized light curve.}
\end{figure}
\begin{figure}
\centerline{\includegraphics[angle=00,scale=.80]{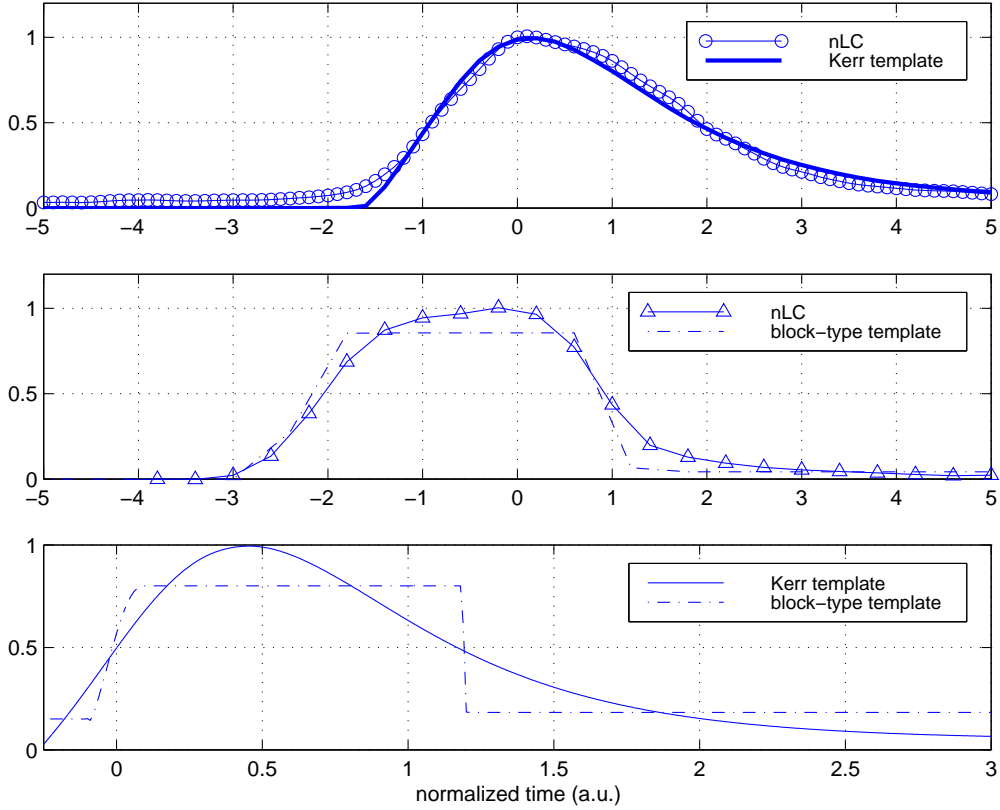}}
\caption{
  Shown are the normalized light curves (nLC) based a Kerr template and a Kerr type
  template for 300 blindly selected long duration events in sample L1: 2 s$<T_{90}<$20 s.
  In each window, the vertical axis refers to normalized count rates (a.u.). The true 
  nLC has the unique property that it matches its template, which applies to the Kerr 
  ({\it top}) but not the block-type template ({\it middle}). Note a slightly smoothed
  average of the block-type templates, due to numerical imperfections in template 
  overlapping. Matching respects the FWHM of the two templates ({\it bottom}).}
\end{figure}
\begin{figure}[b]
\centerline{\includegraphics[scale=.65]{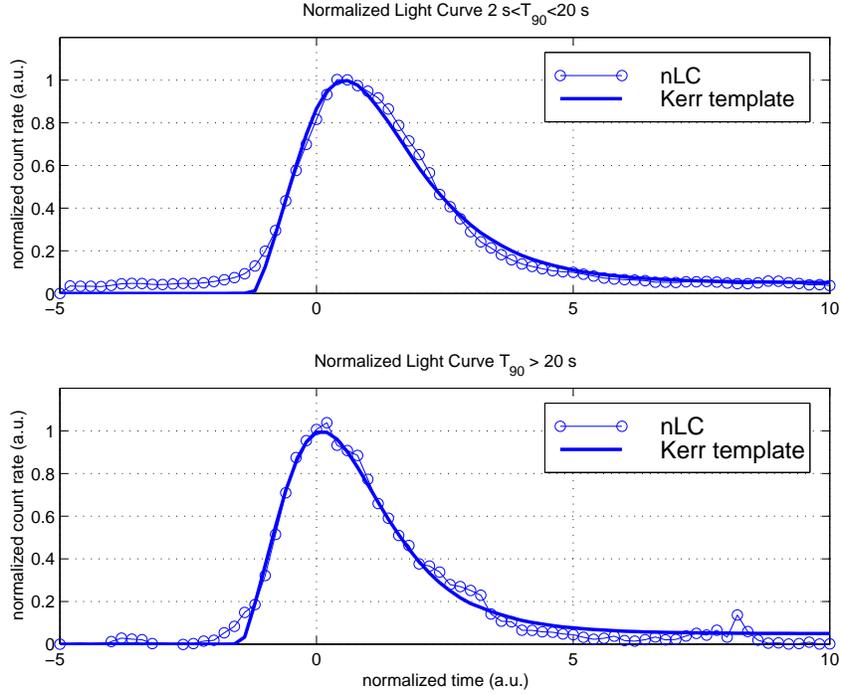}}
\caption{The ensemble average of light curves of long GRBs in the BATSE catalogue is the 
  normalized light curve (nLC), obtained by averaging 300 normalized curves of blindly 
  selected long events in sample L1: 2 s$<T_{90}<$20 s ($top$) and 300 blindly selected 
  long events in L2: $T_{90}>20$ s ($below$). The normalizations are based on matched
  filtering, here against the theoretical template representing viscous spin-down of a
  Kerr black hole. Convergence of the nLC is slower in L2 than in L1, which may be attributed 
  to frequent intermediate periods of quiescence or dead-periods in the actual GRB light curves. 
  The results show excellent agreement between the theory of viscous spin-down, corresponding
  to a nearly doubling of the Bekenstein-Hawking entropy of the even horizon.}
\label{fig:4}    
\end{figure}
\begin{figure}[b]
\centerline{\includegraphics[scale=.45]{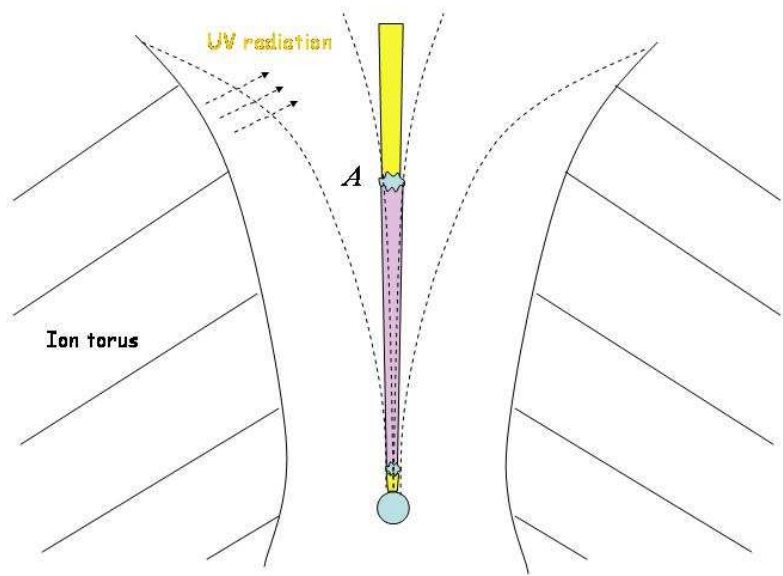}}
\caption{The coupling of the Riemann tensor and angular momentum of charged particles
creates a relativistic capillary effect, whereby pairs are extracted from the vicinity
of the black hole to large distances. The outflow becomes largely force-free up to an
outgoing Alfv\'en front $A$, which mediates a Faraday-induced horizon Fermi-level to large 
distances along equipotential magnetic flux-surfaces.
Upstream of $A$, the magnetic flux-surfaces are essentially
charge free, and stable against pair-cascade at large distances from the black 
hole. (Here, frame-dragging is negligible, as it decays with the cube of the 
distance from the black hole.) As ions are produced from stray particles by 
exposure from UV-radiation from an ion torus, as in the 60 light-year diameter torus 
of M87, they are subject to linear acceleration by the generating voltages at the 
Alfv\'en surface (which are increasing with poloidal angle away from the polar axis).
The linear accelerator here forms in particular to intermittent sources, on a time-scale
less than the crossing time of the radius of the ion torus.}
\label{fig:5}
\end{figure}

\end{document}